\documentstyle[aps,epsf,prl]{revtex}  
\newcommand{\et}{ \mbox{$E_T$}}
\newcommand{\ptran}{\mbox{${P}_{T}$}}
\newcommand{\met}{\mbox{${\not\!\!E_T}$}}
\newcommand{\ttbar}{\mbox{$ {\it t} \bar{{\it t}} $} }

\newcommand{\ppbar}{\mbox{$ {\it p} \bar{{\it p}} $} }

\newcommand{\inp}{\mbox{ ${\rm pb}^{-1}$ } }

\newcommand{\Flong}{\mbox{${\cal F}_{0}$}}

\newcommand{\Fright}{\mbox{${\cal F}_{+}$}}

\newcommand{\gevcc}{\mbox{${\rm GeV}/c^2$}}
\newcommand{\gevc}{\mbox{${\rm GeV}/c$}}
\newcommand{\gev}{\mbox{${\rm GeV}$}}

\newcommand{\W}{\mbox{$W$}}

\def\r#1{\ignorespaces $^{#1}$}

\begin{document}
\title{Measurement of the Helicity of W Bosons in Top Quark Decays}
\draft 


\renewcommand{\baselinestretch}{2}

\author{
\font\eightit=cmti8
\hfilneg
\begin{sloppypar}
\noindent
T.~Affolder,\r {21} H.~Akimoto,\r {42}
A.~Akopian,\r {35} M.~G.~Albrow,\r {10} P.~Amaral,\r 7 S.~R.~Amendolia,\r {31} 
D.~Amidei,\r {24} J.~Antos,\r 1 
G.~Apollinari,\r {35} T.~Arisawa,\r {42}
% ...
\end{sloppypar}
\vskip .026in
\begin{center}
(CDF Collaboration)
\end{center}
\vskip .026in
\begin{center}
\r 1  {\eightit Institute of Physics, Academia Sinica, Taipei, Taiwan 11529, 
Republic of China} \\
\r 2  {\eightit Argonne National Laboratory, Argonne, Illinois 60439} \\
\r 3  {\eightit Istituto Nazionale di Fisica Nucleare, University of Bologna,
I-40127 Bologna, Italy} \\
\r 4  {\eightit Brandeis University, Waltham, Massachusetts 02254} \\
\r 5  {\eightit University of California at Los Angeles, Los 
Angeles, California  90024} \\  
\r 6  {\eightit Instituto de Fisica de Cantabria, University of Cantabria, 
39005 Santander, Spain} \\
\r 7  {\eightit Enrico Fermi Institute, University of Chicago, Chicago, 
Illinois 60637} \\
\r 8  {\eightit Joint Institute for Nuclear Research, RU-141980 Dubna, Russia}
\\
\r 9  {\eightit Duke University, Durham, North Carolina  27708} \\
\r {10}  {\eightit Fermi National Accelerator Laboratory, Batavia, Illinois 
60510} \\
\r {11} {\eightit University of Florida, Gainesville, Florida  32611} \\
\r {12} {\eightit Laboratori Nazionali di Frascati, Istituto Nazionale di Fisica
               Nucleare, I-00044 Frascati, Italy} \\
\r {13} {\eightit University of Geneva, CH-1211 Geneva 4, Switzerland} \\
\r {14} {\eightit Harvard University, Cambridge, Massachusetts 02138} \\
\r {15} {\eightit Hiroshima University, Higashi-Hiroshima 724, Japan} \\
\r {16} {\eightit University of Illinois, Urbana, Illinois 61801} \\
\r {17} {\eightit The Johns Hopkins University, Baltimore, Maryland 21218} \\
\r {18} {\eightit Institut f\"{u}r Experimentelle Kernphysik, 
Universit\"{a}t Karlsruhe, 76128 Karlsruhe, Germany} \\
\r {19} {\eightit Korean Hadron Collider Laboratory: Kyungpook National
University, Taegu 702-701; Seoul National University, Seoul 151-742; and
SungKyunKwan University, Suwon 440-746; Korea} \\
\r {20} {\eightit High Energy Accelerator Research Organization (KEK), Tsukuba, 
Ibaraki 305, Japan} \\
\r {21} {\eightit Ernest Orlando Lawrence Berkeley National Laboratory, 
Berkeley, California 94720} \\
\r {22} {\eightit Massachusetts Institute of Technology, Cambridge,
Massachusetts  02139} \\   
\r {23} {\eightit Institute of Particle Physics: McGill University, Montreal 
H3A 2T8; and University of Toronto, Toronto M5S 1A7; Canada} \\
\r {24} {\eightit University of Michigan, Ann Arbor, Michigan 48109} \\
\r {25} {\eightit Michigan State University, East Lansing, Michigan  48824} \\
\r {26} {\eightit University of New Mexico, Albuquerque, New Mexico 87131} \\
\r {27} {\eightit The Ohio State University, Columbus, Ohio  43210} \\
\r {28} {\eightit Osaka City University, Osaka 588, Japan} \\
\r {29} {\eightit Universita di Padova, Istituto Nazionale di Fisica 
          Nucleare, Sezione di Padova, I-35131 Padova, Italy} \\
\r {30} {\eightit University of Pennsylvania, Philadelphia, 
        Pennsylvania 19104} \\   
\r {31} {\eightit Istituto Nazionale di Fisica Nucleare, University and Scuola
               Normale Superiore of Pisa, I-56100 Pisa, Italy} \\
\r {32} {\eightit University of Pittsburgh, Pittsburgh, Pennsylvania 15260} \\
\r {33} {\eightit Purdue University, West Lafayette, Indiana 47907} \\
\r {34} {\eightit University of Rochester, Rochester, New York 14627} \\
\r {35} {\eightit Rockefeller University, New York, New York 10021} \\
\r {36} {\eightit Rutgers University, Piscataway, New Jersey 08855} \\
\r {37} {\eightit Texas A\&M University, College Station, Texas 77843} \\
\r {38} {\eightit Texas Tech University, Lubbock, Texas 79409} \\
\r {39} {\eightit Istituto Nazionale di Fisica Nucleare, University of Trieste/
Udine, Italy} \\
\r {40} {\eightit University of Tsukuba, Tsukuba, Ibaraki 305, Japan} \\
\r {41} {\eightit Tufts University, Medford, Massachusetts 02155} \\
\r {42} {\eightit Waseda University, Tokyo 169, Japan} \\
\r {43} {\eightit University of Wisconsin, Madison, Wisconsin 53706} \\
\r {44} {\eightit Yale University, New Haven, Connecticut 06520} \\
\end{center}
}
\maketitle
\wideabs{
We use the transverse momentum spectrum of leptons in the decay chain
$ t \rightarrow b W$ with $W \rightarrow l \nu$ to measure the helicity of 
the \W\
bosons in the top quark rest frame. Our measurement uses a \ttbar sample 
isolated in $106 \pm 4$~\inp~of data collected in $\ppbar$ collisions 
at $\sqrt{s} = 1.8$ TeV with the CDF detector at the Fermilab Tevatron.
Assuming a standard V--A weak decay, we 
find that the fraction of \W's with zero helicity in the top
rest frame is $\Flong=0.91 \pm 
0.37 {\rm (stat)} \pm 0.13 {\rm (syst)}$, consistent with the standard 
model prediction of $\Flong = 0.70$ for a top mass of 175 \gevcc.  
\pacs{14.65.Ha, 12.15.Ji, 12.60.Cn, 13.88.+e}

\rightline{FERMILAB-PUB-99/257-E.}
\rightline{September 15, 1999}

}

%
%





The weak decays of the top quark should be described by the universal V--A
charged-current interactions of the standard model. The theory makes a
specific prediction for the polarization state of the \W\ 
bosons, which can be measured using the lepton momentum spectrum in 
the decay  chain $ t \rightarrow b W$ with $W \rightarrow l \nu$. 
Because the top, with mass $m_t = 174.3 \pm 5.1$ \gevcc~\cite{combmass},
is heavier than the \W, 
the \W\ polarization in top decay is fundamentally different from that of other
weak decays. Observation of the predicted lepton momentum spectrum 
can verify that this is the top quark of the standard model. 

In top decays with a pure V--A coupling the amplitude for
positive helicity $\W^+$ bosons is suppressed by
the chiral factors of order $m_b^2/M_W^2$, and the \W\ helicity is
a superposition of just the zero and negative helicity
states~\cite{flip}. At tree level in the standard model, the relative 
fraction \Flong\ of the longitudinal (or zero helicity) \W's 
in the top rest frame is predicted to be~\cite{polemetry}:
\begin{equation}
\label{eq:brdist}
\Flong  = \frac{m_t^2/2 M_W^2}{1+m_t^2/2 M_W^2} 
        = (70.1 \pm 1.6)\%  
\end{equation}
This expression is valid when $m_t$ is significantly greater than
$M_W$.  The
dominance of the zero helicity state may be understood in terms of 
the large top Yukawa coupling to the longitudinal mode of the \W.

We will use \Flong\ to parametrize the agreement
between the predicted and measured lepton momentum spectrum in top decay. 
Effective Lagrangian treatments can be 
used to relate the value of \Flong\ to the strength of non-standard
decay couplings \cite{polemetry,willenbrock}.  Indirect limits 
on such couplings have been derived from precision b quark measurements
\cite{sally,whisnant}. The strictest of these uses the measured rate
of $b \rightarrow s \gamma$ to limit the size of a V+A contribution 
to top decay to less than a few percent~\cite{whisnant,cleo}.
We address the matter of a direct test for a V+A contribution in top 
decay separately at the end of this paper. 

\begin{figure}
\begin{center}
\begin{minipage}[t]{2.9in}      
\hspace{-0.4in}
 \epsfxsize=\textwidth
 \epsfbox[0 162 500 657]{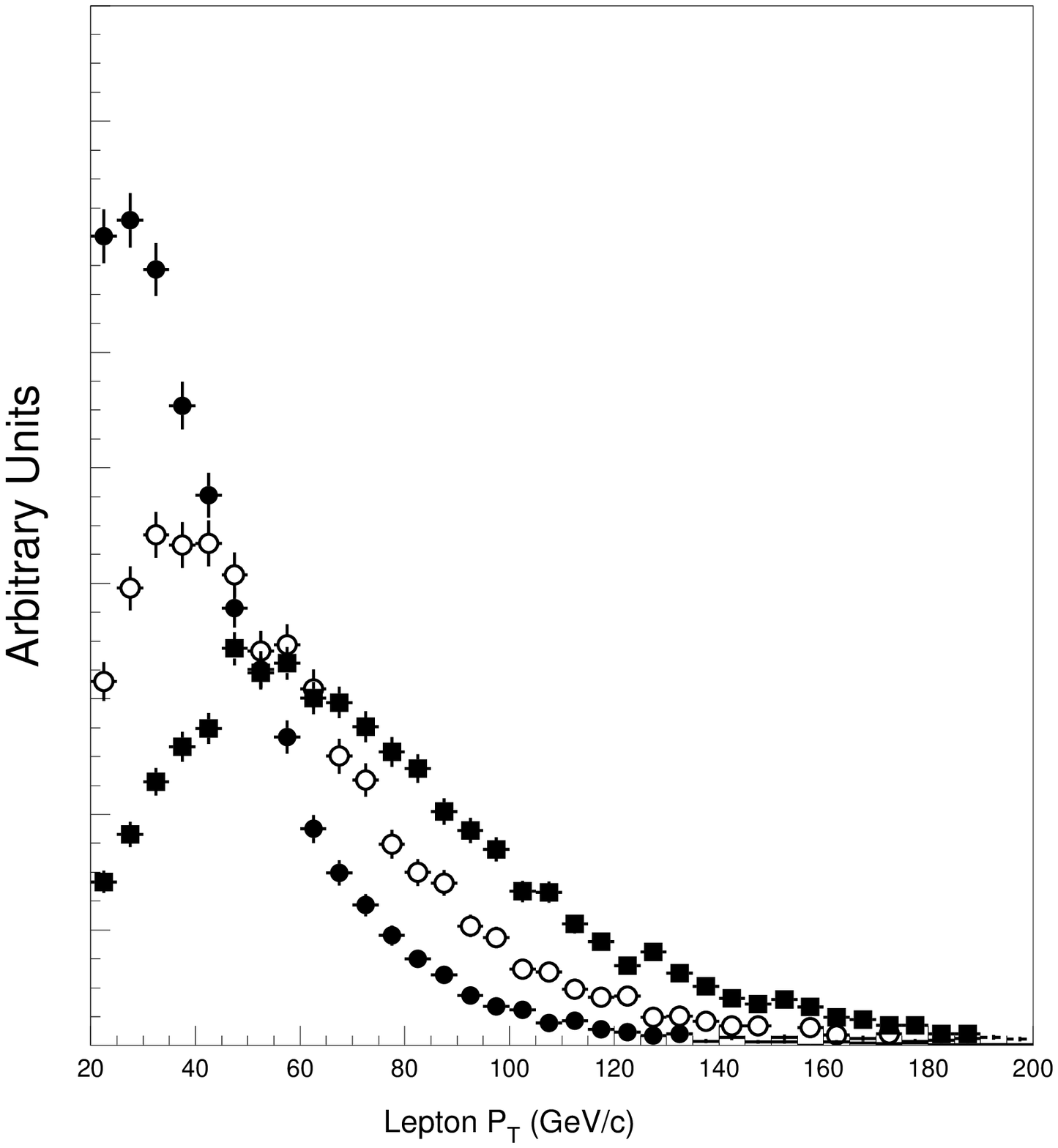}
\vspace{0.1in}
\caption{Lepton \ptran\ distributions for the three \W\ helicities.  The solid
circles are from negative helicity $\W^+$ and positive helicity $W^-$,
the open circles are from longitudinal $\W^+$ and $\W^-$, and the closed
squares are from positive helicity $\W^+$ and negative helicity $\W^-$.  All
three distributions are normalized to the same area.}
\label{fig:model}
\end{minipage}
\end{center}
\end{figure}

We measure \Flong\ in \ttbar\ decays where one or both of the \W's from top 
decays leptonically. 
The V--A coupling at the lepton vertex induces a strong correlation
between the \W\ helicity and lepton momentum which 
survives into the lab  frame.  Charged leptons from negative 
helicity \W\ are softer than the charged leptons from longitudinal \W\ 
bosons.  In Figure~\ref{fig:model} we show the expected
lepton transverse momentum (\ptran) in the laboratory frame~\cite{coord} for 
the 
three \W\ helicities. These spectra are generated from a custom 
version of the HERWIG Monte Carlo program with adjustable \W\ helicity 
amplitudes~\cite{herwig}, followed by a complete simulation of the detector 
effects.  The threshold at 20 \gevc\ is a result of 
our event selection, and will be discussed below.  

To measure \Flong\ we model the lepton \ptran\ in 
$t \rightarrow b l \nu$ according to the standard model as a superposition 
of the \W\ boson negative and zero helicity 
distributions in Figure~\ref{fig:model}, and then use a maximum likelihood
method to find the relative ratio which best fits the data. 
Our measurement uses a \ttbar sample isolated in $106 \pm 4$ \inp 
of data collected in $\ppbar$ collisions at $\sqrt{s} = 1.8$ TeV 
with the CDF detector at the Fermilab Tevatron. The detector is  
described in \cite{det}. 

Decays of \ttbar pairs with a single lepton, 
called lepton+jet events, are characterized by a single
isolated high \ptran\ electron or muon, missing transverse 
energy (\met) from the neutrino in the $\W \rightarrow l \nu$ decay,
and four jets, two from the hadronically decaying \W\ boson and two
from the $b$ quarks.  
Our lepton+jet sample is selected by requiring a single electron or muon
with $\ptran > 20$
\gevc\ which is isolated from jet activity, $\met > 20$ \gev,  and 
at least three jets with measured $\et > 15$ \gev.  

In the manner of previous CDF top analyses, we divide the lepton+jet
events into subsamples based on three selections with different 
top purities.  In the SVX tag 
sample, we require at least one of the jets in the event to be identified as a 
$b$ jet candidate by reconstructing a secondary vertex from the $b$ quark 
decay using the silicon vertex tracker (SVX).  The SVX tagging algorithm is 
described in \cite{tag}.  In the soft lepton tag (SLT) sample, 
we require that one or more jets be identified as a $b$ jet candidate by 
identifying an additional lepton in the event, which is presumed to come from 
a semi-leptonic $b$ decay (see~\cite{tag}).  We also require a fourth jet
in the event which has $\et > 8$ \gev\ and $\mid \eta  \mid < 2.4$.  
Events that satisfy the requirements of both the SVX and SLT samples are
considered to be SVX events, and are removed from the SLT tag sample.
In the No-Tag sample, we require a fourth jet in the event with
$\et > 15$ \gev\ and $\mid \eta \mid < 2.0$.
The backgrounds in the SVX 
sample are described in \cite{xsec}, while those in the No-Tag 
and SLT tag sample 
are given in~\cite{mass}.


%
%
%


Events where both \W's from top decay into leptons,
called dilepton events, are characterized 
by an electron or muon plus \met\ from each of the two $\W \rightarrow l \nu$ 
decays, and two jets from the $b$ quarks.  The two leptons must be oppositely 
charged.  The selection requirements and backgrounds we use for the dilepton
sample are described in~\cite{dil}.  We make the additional requirement that 
the two leptons not be of the same flavor.  This cut removes a background from 
Drell-Yan events with large \met\ for which we have no good lepton \ptran\ 
model.  It removes 2 of the 9 events in the standard CDF dilepton analysis 
(see Ref.~\cite{dil}), but reduces the 
background from $2.4 \pm 0.5$ to $0.76 \pm 0.21$ events, for an overall gain in 
purity.


The largest source of background in the lepton+jet sample 
consists of
\W\ bosons produced with associated jets, called \W+jets events. 
We model these, as well as other smaller contributions,
using VECBOS~\cite{vecbos}, a Monte Carlo 
program that has been shown to be a good representation of these processes
\cite{vecbosproof}.  A smaller, but still significant lepton+jet 
background, 
($23 \pm 5$)\% averaged across the three lepton+jet subsamples, 
comes from non-\W\ events, 
i.e.\ fake leptons and heavy quark production.  We 
use lepton+jet data 
events, in which the lepton is embedded in jet activity and fails our lepton
isolation requirement for the top sample, to model these backgrounds.

The background to the dilepton sample comes from $Z
\rightarrow \tau \tau$, $WW$, $WZ$, and $ZZ$ production, and fake lepton events 
where a jet passes our lepton identification cuts.  We model these backgrounds
using a combination of the PYTHIA and ISAJET Monte Carlo generators
\cite{pythia,isajet} and CDF data~\cite{dil}. 

We summarize in 
Table~\ref{tab:long} the number of events and the predicted amount of
background in each data sample.  Note that the dilepton sample contributes 2 
entries for each event.





We use an unbinned log-likelihood function to estimate the fraction of
top quarks that decay to longitudinal \W\ bosons.  Let ${\cal P}^{S} (\ptran 
; \Flong,m_t)$ be the probability density to obtain a lepton with transverse 
momentum \ptran\ from a top quark of mass $m_t$ and longitudinal fraction
\Flong.  To obtain ${\cal P}^S$ we generate two samples of \ttbar\ events at
mass $m_t$, 
using the HERWIG Monte Carlo generator in concert with a full detector
simulation. In one sample top decays only to negative helicity
\W\ bosons and in the other top decays only to longitudinal 
\W\ bosons.  
We then parameterize the lepton \ptran\ spectrum of each sample as the product 
of an exponential and a polynomial.  
We add the resulting functions together,
using the factors $1 - \Flong$ and \Flong\ as weights for the
respective components. This yields the probability density ${\cal P}^S$ as a 
smooth function of \Flong\ and a discrete function of $m_t$.  
The probability density ${\cal P}^B(\ptran)$ of finding a lepton with 
transverse momentum \ptran\ in the background to our top signal is obtained 
via a similar parameterization of background model lepton \ptran\ 
distributions.  Both ${\cal P}^S$ and ${\cal P}^B$ are normalized to a
probability of 1 above the lepton \ptran\ threshold of 20 \gevc.

The negative log-likelihood is the sum of two terms:
\begin{equation}
\label{eq:like}
-\log{{\cal L}} = -\log{{\cal L}_{\it shape}} - \log{{\cal L}_{\it backgr}},
\end{equation}
where ${\cal L}_{\it shape}(m_t,x_b,\Flong)$ represents the joint probability 
density for a sample of $N$ leptons with transverse momenta $P_{T i}$ to be 
drawn from a population of top candidate events with mass $m_t$, background 
fraction  $x_b$, and longitudinal \W\ fraction \Flong:
\begin{equation}
{\cal L}_{\it shape} = \prod_{i=1}^{N} 
		      [(1-x_b){\cal P}^S(P_{T i};\Flong,m_t)
                         +x_b {\cal P}^{B}(P_{T i})].
\end{equation}
We compute the log-likelihood for each of our analysis subsamples separately,
and then add them together and minimize them simultaneously.
The ${\cal L}_{\it backgr}$     
term in Equation~\ref{eq:like} is included to allow
us to constrain the background fraction $x_b$ to the expected values as
shown in Table~\ref{tab:long}.  
In the lepton+jet subsamples the background estimates are
given as a fraction of the size of the sample, so we
use a Gaussian probability density
$G(x_b,\langle x_b \rangle, \sigma_b)$ with mean $\langle x_b \rangle$ and 
width $\sigma_b$ given by the independent background 
measurements~\cite{xsec,mass} to constrain $x_b$ directly.  
In the dilepton subsample we have an absolute prediction for the number of
background events, so we place a Gaussian constraint on $n_b$, the number of
background events in the sample, with the Gaussian mean and width drawn from
the background study in~\cite{dil}.
We additionally constrain the
sum of the signal and background contributions to the dilepton subsample with 
a Poisson probability density function $P(N,n_s+n_b)$ in $N$ with mean 
$n_b+n_s$, where $N$ is the number of events in the dilepton subsample and 
$n_s$ is the number of signal events in the subsample.  In this case
$N$, $n_b$, and $n_s$ are variable parameters in the log-likelihood 
minimization, and $x_b$ is derived from the relation $x_b = n_b/N$.


The result must be corrected for an acceptance bias caused by the minimum
lepton \ptran\ requirement. Although our Monte Carlo \ptran\ 
distributions account for detection effects on the  shapes of the
lepton \ptran\ distributions we must
separately correct for the difference in efficiency of the \ptran\ cut for
leptons from longitudinal and negative helicity W bosons.
The stiffer longitudinal \W\ decays are 
30\% more likely to be accepted than negative helicity decays. The 
magnitude of the induced bias depends upon the extracted value 
of \Flong; it adds 0.08 to the measured value when the 
true value is near 0.50, but vanishes
as \Flong\ approaches 0 or 1. This correction also modifies the
statistical uncertainty of the measurement.

\begin{figure}
\begin{center}
\begin{minipage}[t]{2.9in}       
\hspace{-0.4in}
 \epsfxsize=\textwidth
 \epsfbox[0 162 500 657]{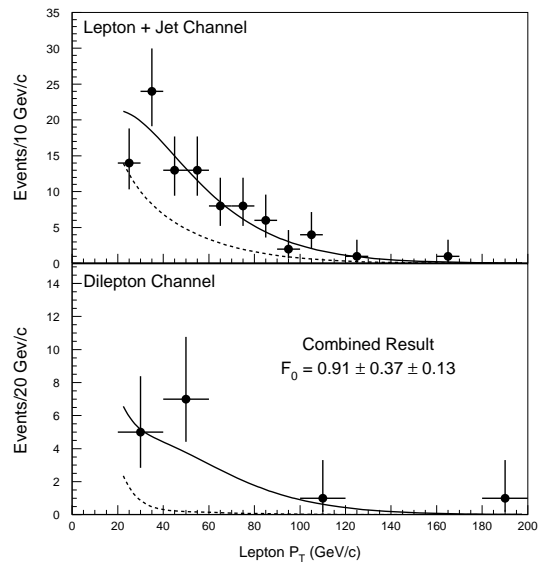}
\vspace{0.1in}
\caption{Lepton \ptran\ distributions for the lepton+jet and dilepton
subsamples.  The lepton+jet subsamples are added together to simplify
presentation.  The data (points) are compared with the result of the combined
fit (solid line) and with the background component of the fit (dashed
line).}
\label{fig:data}
\end{minipage}
\end{center}
\end{figure}

We minimize the log-likelihood with 
respect to \Flong\ at a top mass of 175 \gevcc\ and obtain 
$\Flong = 0.91 \pm 0.37$, after subtracting 0.02 from the result of the
minimization to account for the acceptance bias.
The statistical uncertainty corresponds to a half-unit 
change in the negative log-likelihood with respect to the minimum.  
In Figure~\ref{fig:data} we compare 
${\cal L}_{\it shape}$ to the lepton+jet and dilepton data distributions.  
We summarize the measurement of \Flong\ in Table~\ref{tab:long}.
Included in this table are the results of measurements performed 
separately
in each subsample. Most of the precision comes from the 
lepton+jet 
events that pass the SVX tagging criteria because it is a large sample and 
has a relatively small background.  We have verified in Monte Carlo studies 
that including the less pure No-Tag and SLT events can increase the
precision of our result by 10--15\%.

The systematic uncertainties associated with this measurement of \Flong\
are listed in Table~\ref{tab:sys}. The largest possible error is due to 
the uncertainty on the top quark mass, because the lepton \ptran\ spectrum
depends upon the mass of the top.
The magnitude of the effect is 
estimated by repeating the analysis on Monte Carlo samples where we 
vary the top mass. For 
$\delta M_t = 5.1$ \gevcc, $\delta \Flong\ = 0.07$ \cite{combmass}.


Another significant systematic uncertainty is due to background 
normalization. The lepton \ptran\ spectrum for non-\W\ processes 
peaks at low \ptran, mimicking the shape from negative 
helicity \W\ bosons. The effect on our measurement is 
estimated by varying the amount of non-\W\ contribution in our background
shapes within the envelope of normalization errors.  
We must also account for a 20\% uncertainty in the tagging efficiency of
the SVX algorithm; this causes a $\pm 0.05$ uncertainty in the measurement
of \Flong.
Other sources of uncertainty include the limits on the
generation of Monte Carlo statistics, the acceptance bias introduced by the
selection cut on the transverse momentum of the lepton, the shape of the
non-\W\ background,
the modeling of initial and final state gluon radiation in
our Monte Carlo samples, and the parton distribution functions. Adding 
all of the uncertainties in quadrature, our final result is 
$\Flong = 0.91 \pm 0.37 ({\rm stat}) \pm 0.13 ({\rm syst})$.



Finally, we return to the question of a V+A component in top decay.
Although the indirect limit from $b \rightarrow s \gamma$ is already
severe, we can still, in principle, use our technique to search
directly for a V+A component in the lepton \ptran\ spectrum. 
As shown in Figure~\ref{fig:model}, 
the momentum of leptons from positive helicity $\W^+$ are harder 
than, and distinguishable from, those with negative or longitudinal 
helicity.  We have accordingly generalized our ${\cal L}_{\it shape}$ to 
include the positive helicity fraction \Fright.
Fitting the lepton \ptran\ spectrum for all three components 
simultaneously, we find no statistical sensitivity with our data set.
As an alternative, we hold \Flong\ constant at its standard model 
value, and fit for the superposition of positive and negative 
helicity \W's, yielding a positive helicity fraction 
$\Fright\ = 0.11 \pm 0.15$.
To find an upper limit on \Fright\ we exponentiate the log-likelihood and 
integrate beneath it between $\Fright = 0.0$ and $\Fright=0.30$.  We set a 
95\% confidence level limit such that 95\% of the area under the 
likelihood is to the left of our upper bound.
We find $\Fright < 0.28$.  Note that the assumption that $\Flong = 0.70$
already requires $\Fright \leq 0.30$.

In summary, we have compared the lepton \ptran\ spectrum in  
semileptonic decays $ t \rightarrow b \W \rightarrow b l \nu$ to the 
predictions of the standard electroweak model for top quark decay. Assuming a 
pure V--A coupling, we measure the fraction of longitudinal \W\
bosons in top quark decays to be 
$0.91 \pm 0.37 ({\rm stat}) \pm 0.13 ({\rm syst})$.  This
measurement is consistent with the prediction of 0.70 
for top quarks of mass 174.3 \gevcc. 

We thank the Fermilab staff and the technical staffs of the participating 
institutions for their vital contributions.  This work is supported by the U.S.
Department of Energy and the National Science Foundation, the Natural Sciences
and Engineering Research Council of Canada, the Istituto Nazionale di Fisica
Nucleare of Italy, the Ministry of Education, Science and Culture of Japan, the
National Science Council of the Republic of China, and the A.P. Sloan
Foundation.


%
%




%
%

\begin{table}
\caption{Result of measurements for \Flong\ and description of 
sample content.  The fifth column lists the measurement after a correction 
for an acceptance bias is applied.  Each dilepton event enters twice in the
last row.}
\label{tab:long}
  \begin{tabular}{lcccc} 
  Sample  & Events & Background & \Flong  & Corrected \Flong \\
    \hline
    SVX tagged & 34 & $9.2 \pm 1.2$ & $0.92^{+0.41}_{-0.41}$ & 
                                                $  0.90^{+0.46}_{-0.46}$ \\
    SLT tagged & 14 & $6.0 \pm 1.2$ & $-0.07^{+0.91}_{-0.27}$ & 
                                                $ -0.07^{+0.87}_{-0.27}$  \\
    No tag     & 46 & $25.9 \pm 6.5$ & $1.15^{+0.98}_{-0.70}$ &  
                                                $  1.15^{+0.98}_{-0.77}$  \\
    Dilepton & 7 & $0.76 \pm 0.21$ & $0.60^{+0.57}_{-0.47}$ & 
                                                $  0.56^{+0.57}_{-0.45}$  \\
    \hline
    Total Leptons & 108 & $42.6 \pm 6.7$ & $0.93^{+0.32}_{-0.32}$ & 
                                                $  0.91^{+0.37}_{-0.37}$ \\
  \end{tabular}
\end{table}

\begin{table}
\caption{List of systematic uncertainties in the measurement of the
helicity of \W\ bosons in top decays.}
\label{tab:sys}
  \begin{tabular}{lc} 
    Source & Uncertainty in \Flong  \\ 
\hline
Top Mass Uncertainty      & 0.07 \\ 
Non-\W\ Background Normalization  & 0.06 \\ 
b-tag efficiency          & 0.05 \\ %
Monte Carlo statistics    & 0.05 \\ 
Acceptance Uncertainties  & 0.02 \\ 
Non \W\ background shape    & 0.04 \\ 
Gluon Radiation           & 0.03 \\ 
Parton distribution functions       & 0.02 \\ 
\hline
Total Uncertainty         & 0.13 \\ 
  \end{tabular}
\end{table}


\begin{references}
\bibitem{flip}
Reference to the decay $t \rightarrow bW^{+} \rightarrow bl^{+}\nu$
implicitly includes the CP conjugate decay $\overline{t} \rightarrow 
\overline{b} W^{-} \rightarrow \overline{b} l^{-} \overline{\nu}$, in which 
the $\W^-$ is in a superposition of the positive helicity and longitudinal 
states.
\bibitem{coord}
 In the CDF coordinate system, $\theta$ and $\phi$ are the polar and 
 azimuthal angles, respectively, with respect to the proton beam direction
 ($z$ axis).  The pseudorapidity $\eta$ is defined as 
 $-\ln \tan (\frac{\theta}{2})$.  The transverse momentum of a particle is
 $P_T = P \sin \theta$\@.  The analogous quantity using calorimeter energies,
 defined as $E_T = E \sin \theta$, is called transverse energy.  The missing
 transverse energy, \met, is defined as $-\sum E^i_T \hat{n_i}$, where
 $\hat{n_i}$ are the transverse components of the unit vectors pointing from
 the interaction point to the energy deposition in the calorimeter ($i$ runs
 over the calorimeter cells).
\end{references}
\end{document}